# TLEP: A HIGH-PERFORMANCE CIRCULAR e⁺e⁻ COLLIDER TO STUDY THE HIGGS BOSON

M. Koratzinos, A.P. Blondel, U. Geneva, Switzerland; R. Aleksan, CEA/Saclay, France; O. Brunner, A. Butterworth, P. Janot, E. Jensen, J. Osborne, F. Zimmermann, CERN, Geneva, Switzerland; J. R. Ellis, King's College, London; M. Zanetti, MIT, Cambridge, USA.

*Abstract*

The recent discovery of a light Higgs boson has opened up considerable interest in circular e⁺e⁻ Higgs factories around the world. We report on the progress of the "TLEP" concept since last year. TLEP is an e⁺e⁻ circular collider capable of very high luminosities in a wide centre-of-mass ($E_{CM}$) spectrum from 90 to 350 GeV. TLEP could be housed in a new 80 to 100 km tunnel in the Geneva region [1]. The design can be adapted to different ring circumference (e.g. 'LEP3' in the 27 km LHC tunnel). TLEP is an ideal complementary machine to the LHC thanks to high luminosity, exquisite determination of $E_{CM}$ and the possibility of four interaction points, both for precision measurements of the Higgs boson properties and for precision tests of the closure of the Standard Model from the Z pole to the top threshold.

## MODES OF OPERATION

The main mode of operation of TLEP is at 240GeV operating as a Higgs factory. However, TLEP can reach the ttbar threshold of 350GeV, where the energy loss per turn is 9GeV. Furthermore, it has a huge potential for running with very high luminosity at lower energies, providing vital precision measurements on the closure of the Standard Model and hence is sensitive to the High Energy Frontier.

The 90GeV run will be divided into running at the Z pole with no polarization requirements, performing a Z lineshape measurement with transverse polarization (10% is sufficient) for precise energy determination, and running at the Z pole with maximum longitudinal polarization for asymmetry measurements. TLEP will also run at the WW threshold (160 GeV) with polarization capability.

## DESIGN CONSIDERATIONS

Although the TLEP project is still at an early stage, it benefits considerably in maturity from the experience gained at LEP, PEPPII, KEKB and soon superKEKB.

Operating TLEP with high luminosity implies a rather short beam lifetime (a few minutes), due to unavoidable physics processes of radiative Bhabha interactions with a cross-section of around 200mbarn with little energy dependence. This calls for a topping-up approach where the main ring remains at constant energy and another, so-called accelerator ring constantly tops up replacing the lost particles. We are opting for the accelerator ring feeding the main ring every O(10s). Using the SPS acceleration times the ramp time can be 1.6 s. For a beam lifetime of 15 minutes, we need to fill 1% of the beam every 10 seconds. This figure increases to 10% for 100 s lifetimes.

We have taken the effective bending radius of the new tunnel to be 9 km. The luminosity of such a machine depends linearly on the SR power dissipation. We have used 100MW total (50MW per beam) and designed the rest of the accelerator parameters around this. Another important ingredient for high luminosity is how small $\beta_y^*$ can be made and we are opting for $\beta^*_y$ = 1mm. The beam longitudinal size is around 2mm, leading to an hourglass parameter of around 0.7.

LEP2 operated with a maximum beam-beam parameter $\xi_y^{max}$ of 0.08 at 94.5 GeV without reaching the beam-beam limit, which was estimated to be around 0.115 at 98 GeV per IP for simultaneous interactions at four IPs [2]. The $\xi_y^{max}$ at 45.6GeV was measured to be 0.045 (corresponding to around 0.03 for the TLEP bending radius). No data exist for the $\xi_y^{max}$ of less simultaneous IPs or different ring diameters, but KEKB achieved a figure 2.3 times larger than the limit implied by the LEP data (i.e. 0.07 at 45.6GeV). We are using values in the range 0.07 to 0.10. The LEP extrapolated numbers are in the range 0.03 (for TLEP-Z) to 0.15 (for TLEP-t), with the KEKB extrapolation being 2.3 times higher. We are therefore confident that the quoted values are achievable.

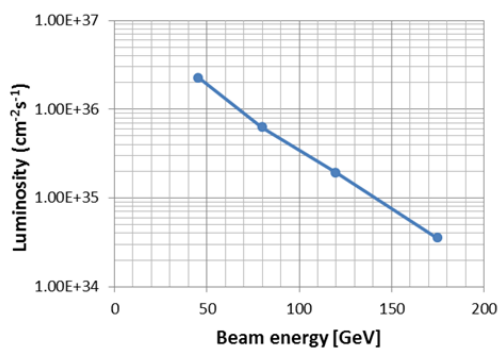

**Figure 1:** TLEP luminosity per IP multiplied by the number of IPs (four)

Here we should note the effect of beamstrahlung, which could have a detrimental effect on integrated luminosity as it might severely affect beam lifetimes [3]. The mitigation options are increasing the ratio of horizontal to vertical emittance ($\kappa_\varepsilon$) – resulting in very flat beams – and the momentum acceptance of the machine. We finally need to ensure that the beamstrahlung-dominated beam lifetime is much larger than the accelerator refill time.

To increase luminosities even further, one needs to resort to more exotic schemes: crab waist collisions, for example, allow for very small $\beta^*_y$ values. Unfortunately, the beamstrahlung problem gets aggravated at the same time, so schemes like charge compensation are called for. We will study the possibilities presented by such exotic schemes to see if they can deliver much higher luminosities and at what cost.

Regarding the emittance ratio $\kappa_\varepsilon$, LEP reached a value of 200, but modern light sources achieve values higher than 2000. We believe that a value of 500 would be attainable with modern diagnostics and a value of 1000 achievable, perhaps also using active magnet supports. An updated list of parameters can be seen in Table 1. The Z pole and WW running parameters are not as advanced as the rest.

We have checked that with the listed parameters a reasonable lifetime can be achieved with momentum acceptance not higher than 2.5%. This represents an improvement over our previous design. For instance, our simulation for TLEP-t gives a beamstrahlung lifetime of 460±50 s for a momentum acceptance of 2.5%.

**Table 1:** TLEP parameters at different energies

|  | TLEP Z | TLEP W | TLEP H | TLEP t |
|---|---|---|---|---|
| $E_{beam}$ [GeV] | 45 | 80 | 120 | 175 |
| circumf. [km] | 80 | 80 | 80 | 80 |
| beam current [mA] | 1180 | 124 | 24.3 | 5.4 |
| #bunches/beam | 4400 | 600 | 80 | 12 |
| #$e-$/beam [$10^{12}$] | 1960 | 200 | 40.8 | 9.0 |
| horiz. emit. [nm] | 30.8 | 9.4 | 9.4 | 10 |
| vert. emit. [nm] | 0.07 | 0.02 | 0.02 | 0.01 |
| bending rad. [km] | 9.0 | 9.0 | 9.0 | 9.0 |
| $\kappa_\varepsilon$ | 440 | 470 | 470 | 1000 |
| mom. c. $\alpha_c$ [$10^{-5}$] | 9.0 | 2.0 | 1.0 | 1.0 |
| $P_{loss,SR}$/beam [MW] | 50 | 50 | 50 | 50 |
| $\beta^*_x$ [m] | 0.5 | 0.5 | 0.5 | 1 |
| $\beta^*_y$ [cm] | 0.1 | 0.1 | 0.1 | 0.1 |
| $\sigma^*_x$ [$\mu$m] | 124 | 78 | 68 | 100 |
| $\sigma^*_y$ [$\mu$m] | 0.27 | 0.14 | 0.14 | 0.10 |
| hourglass $F_{hg}$ | 0.71 | 0.75 | 0.75 | 0.65 |
| $E^{SR}_{loss}$/turn [GeV] | 0.04 | 0.4 | 2.0 | 9.2 |
| $V_{RF}$,tot [GV] | 2 | 2 | 6 | 12 |
| $\delta_{max,RF}$ [%] | 4.0 | 5.5 | 9.4 | 4.9 |
| $\xi_x$/IP | 0.07 | 0.10 | 0.10 | 0.10 |
| $\xi_y$/IP | 0.07 | 0.10 | 0.10 | 0.10 |
| $f_s$ [kHz] | 1.29 | 0.45 | 0.44 | 0.43 |
| $E_{acc}$ [MV/m] | 3 | 3 | 10 | 20 |
| eff. RF length [m] | 600 | 600 | 600 | 600 |
| $f_{RF}$ [MHz] | 700 | 700 | 700 | 700 |
| $\delta^{SR}_{rms}$ [%] | 0.06 | 0.10 | 0.15 | 0.22 |
| $\sigma^{SR}_{z,rms}$ [cm] | 0.19 | 0.22 | 0.17 | 0.25 |
| $\mathcal{L}$/IP[$10^{32}$cm$^{-2}$s$^{-1}$] | 5600 | 1600 | 480 | 130 |
| number of IPs | 4 | 4 | 4 | 4 |
| beam lifet. [min] | 67 | 25 | 16 | 20 |

# POWER CONSUMPTION

For the design study that will follow, it is important to have a preliminary estimate of the power consumption which will serve as the template to improve upon.

## RF system

The RF system is the major power consumer of TLEP which is designed with a maximum SR power dissipation of 100MW (around 1kW per meter of bend). This power is supplied by the RF system whose efficiency is important for the overall power consumption.

The frequency choice is dictated by the compromise of short bunch length for high frequencies on the one hand and limitations in power handling that also increase with frequency on the other. A frequency of 700 to 800 MHz is now considered as a baseline solution. Regarding the accelerating gradient to be chosen, higher gradients result in a more compact RF system, but there is a trade-off with cryogenic power. An accelerating gradient of 20 MeV/m is our baseline.

Regarding power converter efficiency, a typical thyristor 6-pulse power converter for this application has an efficiency of 95%, whereas a switch mode converter runs at 90% efficiency. A klystron run at saturation (as in LEP2) without headroom for RF feedback runs at a 65% efficiency. We consider that fast RF feedback is not necessary for TLEP. RF distribution losses are 5-7%, leading to an overall efficiency of 54%-59%.

To estimate the cryogenic power consumption, we use the LHC figures (900W/W at 1.9 K) to arrive at 23 MW at 175 GeV (fundamental frequency dynamic load only). To this we need to add static heat loads, HOM dissipation in cavities, overhead for cryogenics distribution etc. We estimate that the final power consumption would be 1.5 times the dynamic load consumption, leading to a consumption of 34 MW at 175 GeV.

**Table 2:** Preliminary RF power consumption

|  | TLEP 120 | TLEP 175 |
|---|---|---|
| RF systems | 173-185 MW | |
| cryogenics | 10 MW | 34 MW |
| top-up ring | 3 MW | 5 MW |
| Total RF | 186-198 MW | 212-224 MW |

The accelerator ring should provide the same total accelerating gradient as the main ring, but for much lower beam currents (O(1%) of the total) and for a fraction of the time, depending on its duty cycle (~10%). In any case the RF power budget of the accelerator ring is included in the calculation above, as the total current in both rings is constant. On the other hand, the power of the accelerator ring will be dominated by ramp acceleration and for a 1.6 s ramp length and 155 GeV energy swing, total ramp power is estimated to be 5 MW.

The power requirements of the RF system at 120 and 175 GeV are summarised in Table 2. For lower energies the power will not exceed the values quoted here.

*Rest of the systems*

Regarding ventilation, we extrapolate from the LHC values [4]. The TLEP tunnel will be slightly wider than the LHC tunnel (5.6 m compared to 4 m of LEP, representing a standard European subway tunnel size) and three times longer. Ventilation power for the LHC is 7MW. We scale this by 3 to arrive at a figure of 21 MW.

Regarding cooling, the requirements of TLEP are that 100MW of power in the form of heated water needs to be evacuated from the tunnel. We should note that a fraction of this power (6% in the case of SPS) is taken up by the air in the tunnel (going into our ventilation budget). The power of the cooling system depends on the specific installation. TLEP requires less than $O(10^4)$ m$^3$/hour of water rate. The installed power for a 580 m$^3$/h pump at 7 bar is a motor of 160 kW. To this we should add the power of the motors of the cooling towers which depends strongly on the installation given the environmental impact. The power of the motor in a 10 MW cell of a cooling tower at CERN is 70 kW. Therefore the power needed to evacuate the SR heat for the tunnel is low compared to other power consumers (around 5MW).

**Table 3:** Preliminary TLEP power consumption at 175 GeV

| *Power consumption* | TLEP 175 |
|---|---|
| RF including cryogenics | 224MW |
| cooling | 5MW |
| ventilation | 21MW |
| magnet systems | 14MW |
| general services | 20MW |
| **Total** | ~280MW |

The magnet system power consumption is scaled up from the Large Hadron-electron collider (LHeC) detailed calculations [5] which amounts to 3.6MW. The magnetic field needed for the 60GeV LHeC ring-ring option is similar to the field needed for TLEP at 175GeV, but TLEP needs three times more magnets. The TLEP main ring would consume therefore 11MW and the accelerator ring, assuming a 20% duty cycle, 2MW. The total power consumption of the magnet system (also taking into account a power converter efficiency of 93%) amounts to 14MW at 175GeV, reducing with the square of the energy for lower energies.

General services (lighting, cranes, local control rooms, buildings, etc.) are estimated to be similar to the LHC consumption of 20MW, as the consumption of the experiments, 25MW.

At the Z lineshape scan and longitudinal polarization run, TLEP needs to use polarization wigglers which have a non-negligible power consumption. Our calculations are based on [6] and amount to an additional 12MW of wiggler power.

Power consumption figures of TLEP-175 can be seen in Table 3. This is a first attempt to quantify the power consumption and should be taken as preliminary. Consumption at different energies will not exceed this number.

## POLARIZATION ISSUES

One of the strong points of the TLEP design is the unparalleled beam energy accuracy achievable through transverse beam polarization. Transverse polarization was measured and used at LEP up to 61 GeV per beam, limited by machine imperfections and energy spread [7]. The energy spread scales as $(E_{beam})^2/\sqrt{\rho}$ (where $\rho$ is the bending radius); beam polarization sufficient for energy calibration should therefore be readily available at TLEP up to 81 GeV, i.e. the WW threshold. A new machine with a better handle on the orbit should be able to increase this limit: a full 3D spin tracking simulation of the electron machine of the LHeC project resulted in a 20% polarization at beam energy of 65 GeV for typical machine misalignments [5]. Polarization wigglers would be mandatory for TLEP to decrease the polarization time to an operational value at the Z peak, as without them the polarization time would be nearly 150 hours.

## DESIGN STUDY

We are organising a design study [8] of all aspects of the TLEP project. The conceptual design study will be completed by 2014 and the detailed technical study by 2017, in time for the next European Strategy meeting. An informed decision on the project could then be taken, including the results of the LHC high-energy run.